# Bernoulli correction to viscous losses. Radial flow between two parallel discs

Jordi Armengol[1], Josep Calbó[1], Toni Pujol[2] and Pere Roura[1]

[1]Department of Physics, University of Girona, E17071 Girona, Catalonia, Spain

[2]Department of Mechanical Engineering and Industrial Construction, University of Girona, 17071 Girona, Catalonia, Spain

**Abstract**

For a massless fluid (density = 0), the steady flow along a duct is governed exclusively by viscous losses. In this paper, we show that the velocity profile obtained in this limit can be used to calculate the pressure drop up to the first order in density. This method has been applied to the particular case of a duct, defined by two plane-parallel discs. For this case, the first-order approximation results in a simple analytical solution which has been favorably checked against numerical simulations. Finally, an experiment has been carried out with water flowing between the discs. The experimental results show good agreement with the approximate solution.

## I. Introduction

Introductory physics textbooks for first-level college students usually devote a couple of chapters to fluid statics and dynamics. Regarding the latter, it is common that textbooks present first the behavior of inviscid fluids and then introduce viscosity.[1-3] When dealing with inviscid fluids, two general laws (mass and energy conservation) are used to justify the two equations commonly used to solve problems of flows within ducts: the continuity equation and the so-called Bernoulli equation. The latter, in absence of gravitational effects (that is, if the flow does not have a significant vertical component), is written as:

$$p_0 - p_1 = \frac{1}{2}\rho v_1^2 - \frac{1}{2}\rho v_0^2 \equiv \Delta p_\rho \qquad (1)$$

where $p_0$ and $p_1$ are pressures on sections 0 and 1 respectively, and where the corresponding (uniform) fluid velocities are $v_0$ and $v_1$ (Fig. 1). With subindex "$\rho$" we indicate that the pressure change $\Delta p$ is related to mass density of the fluid, i.e. to its inertia. This equation easily describes the change of pressure along a horizontal duct due to changes in velocity (which, of course, must be related to changes in the duct section).

When viscosity is introduced, the simplest case of a cylindrical duct of constant section is usually analyzed. For this particular geometry, the pressure difference required to balance viscous resistance in the flow, i.e. the so-called Poiseuille equation, is given by:

$$p_0 - p_1 = \frac{8L\eta}{\pi R^4}Q = \frac{8L\eta\bar{v}}{R^2} \equiv \Delta p_\eta \qquad (2)$$

where $\eta$ is the viscosity coefficient of the fluid, $R$ is the duct radius, $L$ is the distance between sections 0 and 1 (where pressures $p_0$ and $p_1$ are applied), and $Q$ is the flow rate. *In contrast with Bernoulli's equation, pressure losses due to viscosity are nonzero even* *for a massless fluid*. Note that, as explained in most textbooks, Poisseuille's equation can be derived from the balance between pressure and viscous forces applied to an infinitesimally thin cylindrical layer of fluid and then integrated over the whole volume (Fig. 2). An intermediate result is the velocity distribution (or profile) within the flow, which turns out to be parabolic:

$$v = \frac{\Delta p_\eta}{4L\eta}\left(R^2 - r^2\right) \qquad (3)$$

This non-uniformity of the velocity across a section of the flow justifies the use of the average velocity ($\bar{v}$) in Eq. 2. Obviously, $\bar{v}$ is defined as $Q/A = Q/\pi R^2$.

A subtle but interesting problem is avoided, however, by most textbooks: what happens when a viscous fluid flows through a non-constant section duct? That is, how should Bernoulli (i.e. inertial) effects be combined with viscous effects? One might think that it is a matter of simply adding the two pressure changes, $\Delta p_\rho$ and $\Delta p_\eta$, and substituting velocities in Eq. (1) by their corresponding average velocities. This is not, however, the exact solution to the problem. An approach to solving this issue has been provided previously in several papers [4-5]. Specifically, one paper[4] shows the generalized Bernoulli equation including transitory and viscous effects, and derives the corresponding specific equations under different conditions (from the simplest steady, incompressible, inviscid flow, to the more complex non-steady and viscous flows). However, equations in that paper are useful only for streamlines (usually for the centerline) and the authors avoid the question of integrating the equation for the whole duct. In another paper[5], an experiment is suggested to demonstrate the importance of considering viscous effects for real fluids (such as water draining out of a cylindrical vessel). In this case, the analysis is so particular that it cannot be applied to other flow situations.

In the present paper, we will approach the question of combining viscous and inertial effects. In Section II we derive a new equation for the first-order approximation to pressure drop in terms of geometry of the duct and of flow rate, which, in turn, depends on both density and viscosity. The general equation derived is then applied to the particular case of a radial flow between two plane-parallel discs. The validity of this approximate solution, i.e. the "first-order Bernoulli correction" to viscous flow, is checked against the results obtained by numerically solving the fluid dynamics (Section III). The approximate solution is then applied to analyze the results obtained with a very illustrative experiment which is described in Section IV. Finally, Section V summarizes the main conclusions of our work.

## II. Theoretical development

### II.a. General expression for pressure change

In general, work associated to pressure changes in a flow is used to 1) change the kinetic energy, 2) change the gravitational energy, and 3) balance the dissipation of energy due to viscosity. If gravitational effects are removed, we can write the relationship with the other two terms, which, expressed as power (i.e. work per unit time) is:

$$(p_0 - p_1)Q = \frac{1}{2}\rho\left[\int_{A_1} v^3 dA - \int_{A_0} v^3 dA\right] + \dot{w}_\eta \tag{4}$$

where $A_0$ and $A_1$ are the areas of sections 0 and 1, respectively, and $\dot{w}_\eta$ is the energy dissipated by viscosity per unit time. The left hand side of Eq. (4) is the power introduced in the system through pressure differences. The first term of the right hand side is the change in kinetic energy, which is often referred to as the Bernoulli term. The other term, as mentioned, is associated with viscosity. Eq. (4) is valid in steady-state conditions if the fluid is incompressible and the velocity of the fluid in any part of sections 0 and 1 is perpendicular to these sections (i.e., the velocity vector is parallel to the differential area vector and pressure is uniform across the section). The latter condition implicitly requires the flow regime to be laminar. Under these conditions, the power introduced by pressure differences can be written as in Eq. (4), since, on a particular section, this work per unit time is

$$\dot{W} = \int_A v\, dF = \int_A v\, p\, dA = p \int_A v\, dA = pQ \tag{5}$$

where $F$ is the force associated with pressure. Regarding kinetic energy, it can be written (for a control volume defined by a given infinitesimal section, $dA$, and the translation of the fluid during a time interval $\Delta t$) as

$$E_c = \frac{1}{2}\int_A \rho (v\,\Delta t\, dA) v^2 = \frac{1}{2}\rho\,\Delta t \int_A (v dA) v^2 = \frac{1}{2}\rho\,\Delta t \int_A v^3 dA\,. \tag{6}$$

Therefore the change of kinetic energy per unit time can be expressed as in Eq. (4). Finally, the power dissipated by viscosity can be calculated through the integral:

$$\dot{w}_\eta = \int_V \eta \dot{\gamma}^2 dV \tag{7}$$

where $V$ means the volume of fluid limited by sections 0 and 1, and

$$\dot{\gamma} = \frac{dv}{dy} \tag{8}$$

is the deformation rate, $y$ being a direction perpendicular to the fluid velocity.

Although Eq. (4) is exact under the conditions mentioned above, its application is not straightforward because it requires the precise knowledge of the velocity at any point of volume $V$. For viscous ($\eta \neq 0$) and dense ($\rho \neq 0$) fluids, analytical solutions for $v$ are difficult to obtain, whereas when $\rho = 0$ or if $v$ is constant along the duct (i.e., there are no changes in section shapes and areas), analytical solutions exist for conduits of simple geometry [6].

Fortunately, we will demonstrate next that Eq. (4) with $\dot{w}_\eta$ calculated from the velocity profile obtained for a massless fluid ($\rho$ = 0) is a good (first-order) approximation to the correct result. Indeed, we can take Taylor's development (on powers of $\rho$) of the main magnitudes given in Eq. (4):

$$
\begin{aligned}
v &= v^{(0)} + \left.\frac{dv}{d\rho}\right|_Q \rho + O(\rho^2) \\
\dot{w}_\eta &= \dot{w}_\eta^{(0)} + \left.\frac{d\dot{w}_\eta}{d\rho}\right|_Q \rho + O(\rho^2)
\end{aligned}
\tag{9}
$$

where superscript (0) means the value computed for $\rho = 0$. Obviously, the zeroth-order approximation to Eq. (4) is:

$$
\Delta p^{(0)} Q = \dot{w}_\eta^{(0)} \tag{10}
$$

since the Bernoulli term is null at the zeroth order. At first order, the Bernoulli term is simply $\frac{1}{2}\rho\left[\int_{A_1} v^{(0)^3} dA - \int_{A_0} v^{(0)^3} dA\right]$ where the zeroth-order approximation to the velocity field can be easily calculated for many simple geometries (the most well-known being the already mentioned parabolic distribution of velocities). Finally, the key point is that the first order approximation of the viscous term is null, that is:

$$
\left.\frac{d\dot{w}_\eta}{d\rho}\right|_Q = 0 \tag{11}
$$

which means that the viscous term is an extremum (minimum) for the distribution of velocities of a massless fluid. A demonstration of this well-known result in fluid dynamics[7] is given in the Appendix for the very simple case of a plane-parallel duct.

Summing up, we can write the first-order approximation to pressure change in a viscous flow within a changing section duct as:

$$\Delta p^{(1)} = \frac{1}{Q}\frac{1}{2}\rho\left[\int_{A_1} v^{(0)^3} dA - \int_{A_0} v^{(0)^3} dA\right] + \frac{\dot{w}_\eta^{(0)}}{Q} \qquad (12)$$

In the next subsection, this expression will be applied to obtain this pressure change in the particular case of a duct defined by two plane-parallel discs where the fluid flows in a radial direction.

### II.b. Viscous flow within two plane-parallel discs

If we have a viscous fluid flowing within a plane-parallel duct with height, *H*, much smaller than its width, *W* (Fig.3), the velocity profile will be parabolic. This can be easily deduced from the balance between pressure and viscous forces, when the section crossed by the flow is constant. The parabolic profile may be written as:

$$v = \frac{\Delta p}{2L\eta}\left(\frac{H^2}{4} - y^2\right). \qquad (13)$$

When this expression is integrated over the whole section, $A = W{\cdot}H$, we obtain:

$$\Delta p = \frac{12L\eta Q}{WH^3}. \qquad (14)$$

Moreover, Eq. (14) can be used in combination with Eq. (13) to write the velocity profile as a function of the flow rate itself:

$$v = \frac{6Q}{WH^3}\left(\frac{H^2}{4} - y^2\right) \qquad (15)$$

Eqs. (13) and (15) are exact for a duct of constant section. They still remain exact for a massless fluid ($\rho = 0$) when *W* changes along the duct. This is so despite the fact that the continuity equation implies a change in velocity, since the kinetic energy is always zero and does not affect the balance of work performed by pressure.

Consequently, these expressions are the *zeroth order* approximation to the velocity distribution between two parallel planes even if the section is not constant.

In our experiment (see Section IV) the fluid flows between two parallel discs in a radial direction (Fig. 4). Therefore, it is a case of a plane-parallel duct where section crossed by the flow is not constant. For the infinitesimal volume between $r$ and $r + dr$, however, Eq. (13) applies, so the velocity profile can be written as follows:

$$v^{(0)} = \frac{1}{2\eta}\left(\frac{H^2}{4} - y^2\right)\frac{dp_\eta}{dr} \tag{16}$$

which, again, when introduced in Eq. (7) and using $dp_\eta = d\dot{w}_\eta / Q$ results in:

$$\frac{\dot{w}_\eta^{(0)}(r)}{Q} = p(R) + \frac{6Q\eta}{\pi H^3}\ln\left(\frac{R}{r}\right) \tag{17}$$

where *p(R)* is the pressure at the outer boundary of the discs, which have $R$ as their maximum radius. This result can also be obtained from Eq. (15), changing $W$ by $2\pi r$, and combining Eqs. (4, 7, and 8).

As far as the Bernoulli term is concerned, it is convenient to use the zeroth-order approximation to the velocity profile as expressed in Eq. (15) and adequately written for our geometry (i.e. $W = 2\pi r$, $dA = 2\pi r dy$). With this, integration of the Bernoulli term according to Eq. (12) gives:

$$\frac{1}{Q}\frac{1}{2}\rho\left[\int_{A_e} v^{(0)3} dA - \int_{A} v^{(0)3} dA\right] = \frac{27\rho Q^2}{140\pi^2 H^2}\left[\frac{1}{R^2} - \frac{1}{r^2}\right]. \tag{18}$$

Finally, the combination of Eq. (17) and Eq. (18) when introduced in Eq. (12) gives the expression that approximates (to the first order) the pressure in a radial flow between two plane-parallel discs:

$$p^{(1)}(r) = p(R) + \frac{27\rho Q^2}{140\pi^2 H^2}\left[\frac{1}{R^2} - \frac{1}{r^2}\right] + \frac{6Q\eta}{\pi H^3}\ln\left(\frac{R}{r}\right) \tag{19}$$

which is written in terms of known or measurable quantities. The second term on the right hand side of the equation is the Bernoulli (or inertial) correction to the pressure drop computed by considering only the viscous effects (i.e., the third term). The validity of this expression will be demonstrated in the following sections.

Before leaving this section, the reader is encouraged to write Eq. (18) in terms of the average velocities, $\bar{v}_1 = Q/2\pi rH$ and $\bar{v}_0 = Q/2\pi RH$, i.e.:

$$\Delta p_{\rho}^{(0)} = \varepsilon\left(\frac{1}{2}\rho\bar{v}_1^2 - \frac{1}{2}\rho\bar{v}_0^2\right) \tag{20}$$

Note that this expression is identical to Bernoulli's elementary Eq. (1) except for the extra factor $\varepsilon$ which takes into account the effect the velocity profile due to viscosity forces. The $\varepsilon$ values depend on the duct section: for a plane-parallel duct $\varepsilon$ = 54/35, whereas $\varepsilon$ = 2 for a cylindrical duct. In spite of the simplicity of Eq. (20) we should remember that this simple generalization of Eq. (1) is not exact.

## III. Numerical simulations

### III.a. The pressure change

Following the experimental design detailed in the next section, here we simulate the flow of water within two plane-parallel discs of radius 20 cm and separation $H$ (= 0.5 mm and 0.25 mm). Several simulations have been carried out with constant values for the volumetric flow rate $Q$ (2, 3, 4 and 5 L min$^{-1}$). In all cases, the inlet corresponds to a drilled hole of radius 1 cm centered on one of the two discs.

It is illustrative to investigate the flow regime in terms of the Reynolds number Re defined as Re = $\rho\bar{v}D/\eta$, where $\bar{v}$ is the average velocity and $D$ is the characteristic length of the problem. In fluid dynamics, Re is commonly used as a control parameter to determine the flow regime (i.e., either laminar or turbulent) in incompressible viscous fluids with negligible external forces. For circular pipe flows, the minimum critical Reynolds number $\mathrm{Re}_c$ is approximately 2000, which means that flow regimes with Re < $\mathrm{Re}_c$ are laminar. We assume a similar value of $\mathrm{Re}_c$ for our case.

For two plane-parallel discs, $D = H$ (see, for instance, ref.[7]) and, from the text above Eq. (20), the average velocity is $\bar{v} = Q/2\pi r H$. Then, the Reynolds number expressed in terms of the flow rate $Q$ reads

$$\mathrm{Re} = \frac{\rho Q}{2\pi r \eta} \tag{21}$$

where, for our experiment, $\rho$ = 1000 kg/m$^3$ and $\eta$ = 1.024 10$^{-3}$ Pa·s (the viscosity of water at 20ºC). Note that Re decreases as r increases. Consequently, Re reaches its maximum value for the radius of the drilled hole (r = 1 cm). At this point, and for the maximum rate of Q = 5 L/min, Re = 1300 which lies well below the critical value $\mathrm{Re}_c$. The laminar regime is thus ensured for r > 1 cm, in consistency with the assumption made in Section II. The type of flow regime is needed in order to select the physical model used by the Computational Fluid Dynamics (CFD) code.

The simulations have been done with a CFD commercial software package based on the finite volume method (STAR-CCM+). The symmetry of the problem allows us to simplify the simulation by analyzing the flow of a two-dimensional (2D) rectangular slice of height $H$ ranging from $r = 0$ to $r = R$ and by imposing an axisymmetric boundary condition for the $r = 0$ edge. This 2D simulation saves computational resources while accepting a high number of cells (i.e., small surfaces where the governing differential equations will be applied). We have divided the rectangular domain into 40,000 cells. Thus, we have squares of side 0.05 mm for the $H$ = 0.5 mm case and rectangles 0.025 mm high and 0.05 mm wide (radial direction) for the $H$ = 0.25 mm case.

In order to reach the steady state, the CFD code ran the number of iterations required for satisfying the convergence criterion based on reaching a threshold value of $10^{-3}$ for the residuals (i.e., weighted differences of the variables between two consecutive iterations). In addition, a test with a 2D model containing over 280,000 cells was performed for the $Q$ = 5 L $min^{-1}$, H = 0.5 mm case and the solution coincided with that obtained with 40,000 cells.

The relative pressure as a function of the radial direction is shown in Fig. 5a for the $H$ = 0.25 mm case by using the theoretical expression (lines) and by numerically solving the fluid flow by means of the CFD code (symbols) for different values of the flow rate. In Fig. 5a, thin solid lines correspond to the analytical solution obtained by neglecting the Bernoulli correction (i.e., Eq. (19) for a massless fluid $\rho$ = 0). On the other hand, thick solid lines refer to Eq. (19) (i.e., including the Bernoulli correction). In both cases *p(R)* has been assumed to be equal to zero, which is the same boundary condition adopted for the pressure outlet in the numerical simulations. From Fig. 5b, we observe that the analytical solution obtained in Eq. (19) reproduces our numerical simulations very well. In contrast, the solution without the Bernoulli correction clearly fails to reproduce the simulations, notably for low values of r. Similar results are obtained for the $H$ = 0.5 mm case (Fig. 5b).

### *III.b Accuracy analysis*

*Despite the good agreement between the analytical and numerical results, one may wonder if, for small radii, this agreement is somewhat fortuitous. The puzzling fact is that although the Bernoulli correction is calculated through the perturbative method developed in Section II.a, it approaches the correct result even when the Bernoulli term is as large as one-half of the viscous term (for H = 0.5 mm, Q = 5 L/min and r = 20 mm*

*in Fig. 5b). Let us have a closer look at the value of the Bernoulli term relative to the viscous term.*

*In fact, the (natural) boundary condition, p = 0 at the outer radius, r = R, implies that the Bernoulli correction is zero at this point and that its value monotonically increases as r gets smaller. Consequently, the Bernoulli term (as well as the viscous one) at any radius $r_0$ is the integration from r = R to $r_0$ of the local pressure increments (dp/dr)·dr. This means that the accuracy of our analytical solution can be better analyzed at the local level by looking at the pressure derivative instead of the pressure itself. This has been done in Fig. 6, where the continuous curves and discrete points correspond to the analytical and numerical solutions, respectively. The analytical curves indicate that Bernoulli's contribution to pressure change is greater than the viscous losses for r < 30 mm. At r = 35 mm (below this radius the numerical simulation becomes unstable for H = 0.5 mm and Q = 5 L/min), the Bernoulli contribution relative to the viscous one is as high as 0.8. Despite this large value, the discrepancy with the numerical result is only 17% of Bernoulli's contribution.*

*The success of the analytical solution in predicting so large corrections when density changes from zero to a finite value is not fortuitous because the Bernoulli term of Eq. (19) is obtained with the hypothesis that the velocity profile almost coincides with that of ρ = 0 (Eq. (12)) (i.e. a parabola) and this is really the case. In the inset of Fig. 6 we have plotted the velocity profile averaged over fluid layers of 0.05 mm. At r = 35 mm, the simulated profile deviates slightly from a parabola, what results in a Bernoulli term 8% larger than the analytical correction. This means that the17% discrepancy in dp/dr is equally shared by the Bernoulli term and the second order term in ρ of the viscous losses (eq.(9)). At r = 60 mm, the velocity profile almost coincides with a parabola (inset of Fig.6), the inaccuracy of the Bernoulli term being less than 1%.*

*The main conclusion reached with the reported numerical simulations is that, within the range of Q's and thicknesses covered by the experiment detailed below, the approximate analytical solution of Eq.(19) is accurate enough to analyze the experimental results.*

## IV. Experiment: design and results

Two discs of radius 20 cm were cut from a 2 cm thick aluminum sheet. One side of the sheet was very smooth and flat, so that deviations from perfect flatness were below ±0.03 mm over the entire disc surface. In the center of one disc a hole of 2 cm diameter was drilled to allow water to flow through, whereas five smaller holes of 0.5 cm diameter were drilled in the other disc at 2, 3, 7, 11, and 15 cm from the center. A plastic hose was connected to every small hole to measure the water pressure by simply quoting the height of the water column inside (see Fig. 7). Although we attempted to measure the pressure near the center and near the disc boundary, we realized that, due to the abrupt variation in flow conditions there, it was not easy to interpret the pressure values at these points.

Between both discs, a thin duct was arranged by positioning them with the help of three pairs of screws. Distance $H$ was fixed with calibrated thin steel sheets of 0.25, 0.30, 0.40, and 0.50 mm. Experiments for thinner ducts were not accurate enough because of the flatness inaccuracy and because the hydrostatic pressure deformed the discs.

A controlled flux rate of water entered the plane duct from below the center of one disc and came out at the disc contour. A floating-ball flow meter was used to select the desired rate $Q$. The chosen values of $Q$ = 2, 3, 4 and 5 L min$^{-1}$ were verified by volumetric measurements. Water temperature remained almost constant at about 20ºC for all the experiments. For every duct thickness, the series of measurements at the four flux rates were repeated four times. In the corresponding figures we quote the average values.

Thus, in Figures 8a-b, we show results of our measurements for the two extreme cases ($H$ = 0.25 mm and $H$ = 0.5 mm respectively), along with the theoretical lines that include the Bernoulli correction (i.e. Eq. 19). Note that experimental values for $r$ = 2 cm are not represented, since at this short distance to the inlet, the flow was hardly stabilized (it could be slightly turbulent as a result of the sharp change in direction at the inlet) producing quite inaccurate measurements. All values shown are not directly the average of the measurements; instead, a correction has been applied in order to take into account the unknown pressure at the outlet edge of the discs (i.e. term $p(R)$ in Eq. 19). Specifically, a pressure correction of few thousands of Pa has been subtracted for each measurement in a series (that is for a given $H$ and $Q$) in such a way that the root mean square difference between the corrected values at $r$ = 7, 11, 15 cm and the corresponding analytical values is minimum.

Figure 8b clearly shows that the experiment that we have carried out is suitable to demonstrate the importance of the Bernoulli correction to viscous effects. Indeed, experimental values match the analytical lines quite well, and would be far from the lines without the Bernoulli correction. This is particularly true for the points at lower $r$, where, as predicted by Eq. 19, the Bernoulli correction has greater effect. Figure 8a also confirms these results, although in this case, the measurements are clearly and systematically lower than the theoretical values, for $r$ = 3 and 7 cm. We think that this error is due to a slight non-flatness of the disks that we caused when using even lower separations ($H \leq 0.2$ mm) between disks. For these extremely narrow flows, numerical simulations have shown that the high static pressure values induced a plastic deformation of the disks.

Another way to present results of our experiments is used in Figure 9a-b. In this figure, the importance of the viscous term in Eq. 19 is stressed, since we show the pressure at a given radius ($r$ = 7 cm) as a function of the other two variables, the separation between disks $H$ and the flow rate $Q$. Figure 9a shows values, for two different $Q$, as a function of $H$. In this case, the dependency between $p$ and $H$ is clearly inverse cubic, as is demonstrated by the reference line with slope -3. Similarly, on the right panel (Fig. 9b) the values of pressure are represented, for two different $H$, as a function of $Q$. In this case, the linear relationship is also clear and it is shown through a reference line of slope 1.

*The experiment here described is complementary to the experimental demonstration of Bernoulli levitation reported in ref. 8. In their experiment, Waltham et al. used air instead of water and the flow was turbulent. The Bernoulli term (Eq. (1)) was higher than the viscous one, what resulted in a negative pressure between the discs.*

## V. Conclusions

Although analytical solutions do not exist for the steady flow of dense fluids along ducts of variable sections, it has been shown that the solution in the limit of $\rho = 0$ can be used to calculate the extra pressure variations due to inertial effects (Bernoulli's correction). This method is correct up to the first order in $\rho$, because the energy dissipation is minimal for the velocity distribution of a massless fluid.

It is possible to illustrate the method with a good experiment, suitable for undergraduate students, consisting of the radial flow of water between two plane-parallel metallic discs separated by a small distance. If elastic and plastic deformations are avoided, then good agreement between theory and experiment can be expected for a reasonable range of experimental conditions.

## Appendix

Here we show that the energy dissipation is minimum for a massless fluid (i.e., $\rho = 0$) in a plane-parallel duct of constant section under the assumption of constant flow rate *Q*.

Let us define *L* as the length of the duct ($0 \le x \le L$), *H* as its height ($-H/2 \le y \le H/2$), and *W* as its width ($0 \le z \le W$, with $W >> H$) (Fig. 3). For symmetry, the velocity vector follows ($u(y)$, 0, 0) and its profile satisfies the Navier-Stokes equation for a massless fluid,

$$0 = -\frac{\partial p}{\partial x} + \frac{\partial \tau_{xy}}{\partial y}, \qquad \text{(A1)}$$

where $\tau_{xy}$ is the viscous stress (i.e., the viscous force on direction *x* acting on a fluid surface normal to direction *y*). For an incompressible fluid in the plane-parallel duct here analyzed, the viscous stress reads,

$$\tau_{xy} = \eta \frac{du}{dy}. \qquad \text{(A2)}$$

By substituting Eq. (A2) into Eq. (A1) we obtain,

$$\frac{d^2 u}{dy^2} = \frac{1}{\eta} \frac{\Delta p}{L}, \qquad \text{(A3)}$$

where $\Delta p/L$ is the constant pressure loss per unit length. The integration of Eq. (A3) leads to the well-known parabolic velocity profile.

The energy dissipation rate due to viscous effects $\delta \dot{w}_\eta$ in our control volume $V$ reads,

$$\dot{w}_\eta = \int_L dx \int_W dz \int_{-H/2}^{H/2} dy \left( \frac{du}{dy} \right) \tau_{xy} . \tag{A4}$$

By substituting Eq. (A2) into Eq. (A4), the energy dissipation rate due to viscous effects is a positive quadratic form,

$$\dot{w}_\eta^{(0)} = \int_L dx \int_W dz \int_{-H/2}^{H/2} dy \eta \left( \frac{du}{dy} \right)^2 , \tag{A5}$$

where the superscript (0) means that this dissipation value corresponds to the velocity profile of the massless fluid satisfying Eq.(13).

Now, let us analyze the implications of slightly modifying this velocity profile. This means that now the velocity at a point is $u + \delta u$, with $\delta u$ *being* the arbitrary perturbation in the velocity field. The non-slip condition implies that this perturbation is zero at the rigid boundaries (i.e., $\delta u(\pm H/2) = 0$). In addition, the condition of constant flow rate $Q$ implies that,

$$\delta Q = \int_W dz \int_{-H/2}^{H/2} dy \delta u = W \int_{-H/2}^{H/2} dy \delta u = 0 . \tag{A6}$$

From Eq. (A4), the energy dissipation rate, $\dot{w}_\eta$, corresponding to the perturbed velocity profile reads,

$$\dot{w}_\eta = \int_L dx \int_W dz \int_{-H/2}^{H/2} dy \eta \left[ \frac{d(u + \delta u)}{dy} \right]^2 = LW\eta \int_{-H/2}^{H/2} dy \left[ \left( \frac{du}{dy} \right)^2 + 2 \frac{du}{dy} \frac{d\delta u}{dy} + O(\delta u^2) \right] . \tag{A7}$$

Note that the contribution of the first term within the square brackets in the last equality in Eq. (A7) corresponds to $\dot{w}_\eta^{(0)}$ (see, Eq. (A5)), whereas the second term refers to the first-order perturbation of the energy dissipation rate $\delta \dot{w}_\eta$. Since, from Eq. (A3), $du/dy \propto y$, we have that,

$$\delta \dot{w}_\eta \propto \int_{-H/2}^{H/2} dy\, y \frac{d\delta u}{dy} = [y\delta u]_{-H/2}^{H/2} - \int_{-H/2}^{H/2} dy \delta u = 0\,, \qquad \text{(A8)}$$

where the first integral has been solved by parts and the last equality follows from the boundary conditions at the rigid boundaries (first term) and from Eq. (A6) (second term).

The condition here found that $\delta \dot{w}_\eta = 0$ for a velocity profile corresponding to a massless fluid subject to the constraint of constant flow rate, is equivalent to find the extremum of $\dot{w}_\eta$ as expressed in Eq. (9). Since the energy dissipation rate $\dot{w}_\eta$ for a Newtonian incompressible fluid is a positive quadratic form, it follows that the extremum is a minimum.

We propose that readers derive the condition $\delta \dot{w}_\eta = 0$ for a duct of constant arbitrary section. If they follow a procedure similar to that given in this Appendix, they will find a general property of the solution of the Navier-Stokes equation useful, namely:

$$\left(\frac{\partial u}{\partial x_i}\right)^2 = u \frac{\partial^2 u}{\partial x_i{}^2}\,, \qquad x_i = y, z. \qquad \text{(A.9)}$$

**References**

1. P. A. Tipler, *Physics for Scientists and Engineers*, (W. H. Freeman and Company / Worth Publishers, New York, 1998) 4th ed., vol.1..
2. H. D. Young and R. A. Freedman *University Physics*, (Addison Wesley Longman, USA, 1996) 9th ed..

3. R. A. Serway, *Physics for Scientists and Engineers with Modern Physics*, (Saunders College Publishing, 1986). 4th ed.

4. C. E. Synolakis and H. S. Badeer, "On combining the Bernoulli and Poiseuille equation - A plea to authors of college physics texts", Am. J. Phys. **57**(11), 1013-1019 (1989).

5. M. E. Saleta, D. Tobia and S. Gil, "Experimental study of Bernoulli's equation with losses", Am. J. Phys. **73**(7), 598-602 (2005).

6. J. Lekner, "Viscous flow through pipes of various cross-sections", Eur.J.Phys. **28**, 521-527 (2007).

7. C. R.. Doering and P. Constantin," Variational bounds on energy dissipation in incompressible flows: Shear flow", Phys.Rev.E **49** (5), 4087-4099 (1994).

8. C. Waltham, S. Bendall and A. Kotlicki, "Bernoulli levitation", Am.J.Phys. **71** (2), 176-179 (2003).

**Figure captions**

Figure 1. Flow of an inviscid fluid through a horizontal duct of changing section.

Figure 2. Viscous forces in a cylindrical duct.

Figure 3. Duct defined by two plane parallel surfaces held at a short distance W.

Figure 4. Radial flow between two parallel discs.

Figure 5a-b. Relative pressure as a function of the radial direction for both $H = 0.25$ mm (left) and $H = 0.5$ mm (right) cases. Thick lines correspond to the analytical solution including the Bernoulli correction for different values of the flow rate. In contrast, thin lines refer to the analytical solution without taking the Bernoulli correction into account. Symbols show the results obtained through numerical simulations.

*Figure 6. Deviation of the pressure change for H = 0.5 mm and Q = 5 L/min. The numerical simulation (open circles) becomes unstable below r = 0.35 mm for this particular H and Q values. Inset: simulated velocity profile (empty circles) compared to the parabolic profile (full circles) at 35 (a) and 60 mm (b).*

Figure 7. A diagram of the experimental set-up.

Figure 8a-b. Relative pressure as a function of the radial direction for both $H = 0.25$ mm (left) and $H = 0.5$ mm (right) cases. Lines correspond to the analytical solution

including or not the Bernoulli correction for different values of the flow rate. Symbols show the results obtained through measurements. *The error bars are the standard deviation of 8 (for r > 50 mm) or 4 (r = 30 mm) measurements.*

Figure 9a-b. (Left) Relative pressure as a function of separation between disks, at $r$ = 7 cm and for two different flow rates. (Right) Relative pressure as a function of flow rate, at $r$ = 7 cm and for two different separations between disks. Reference lines have slopes -3 and 1 respectively.

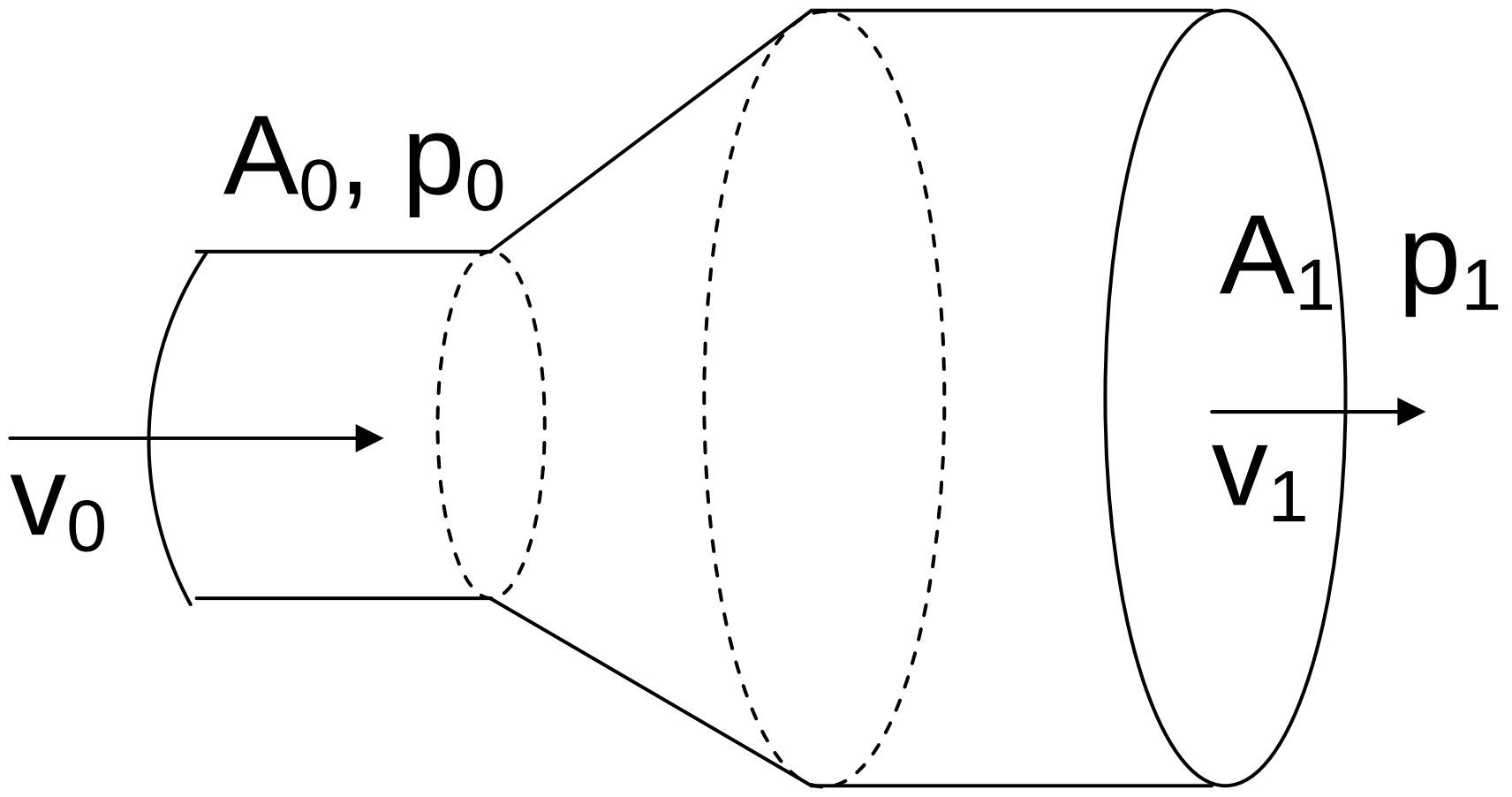

Fig.1.-

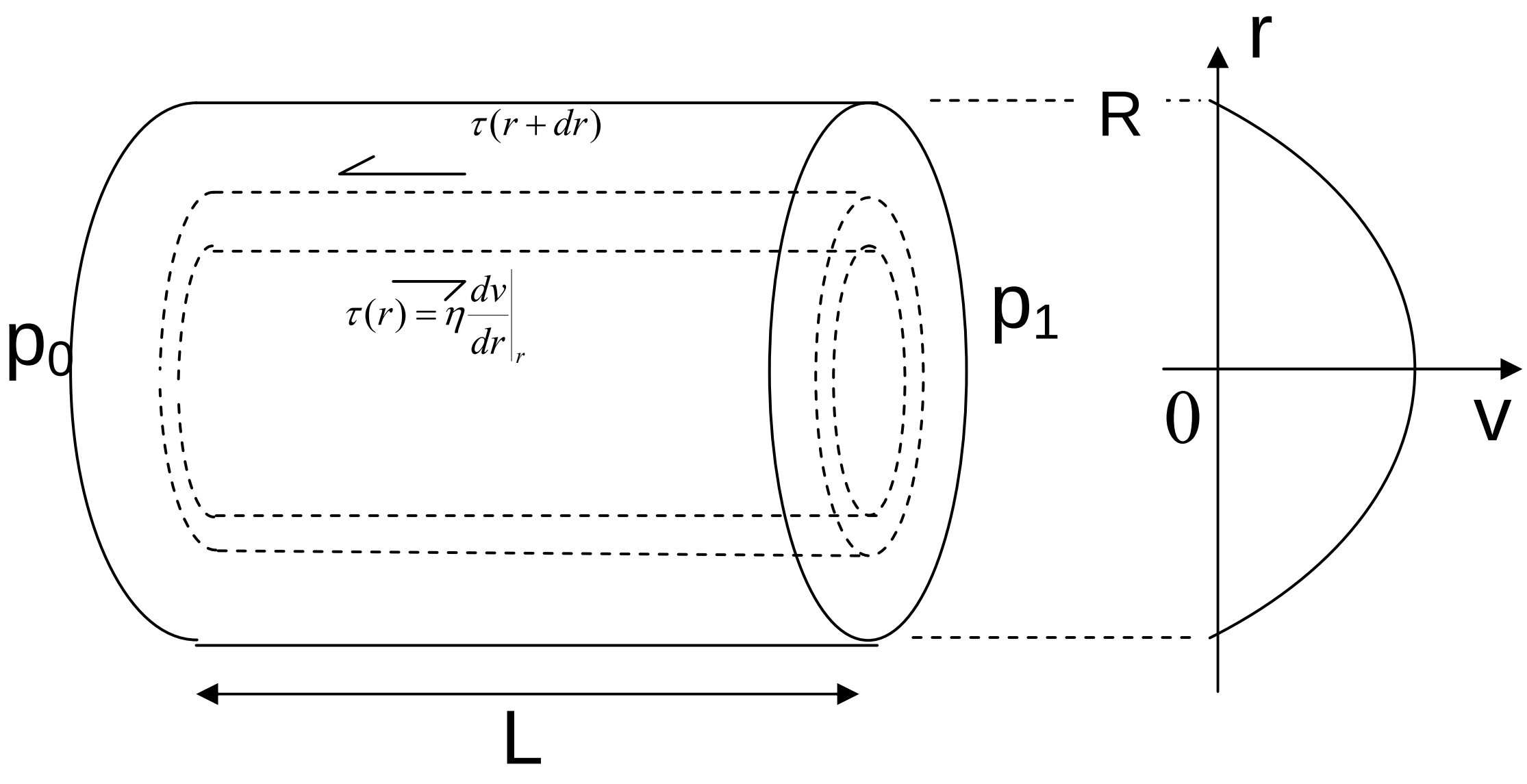

Fig.2.-

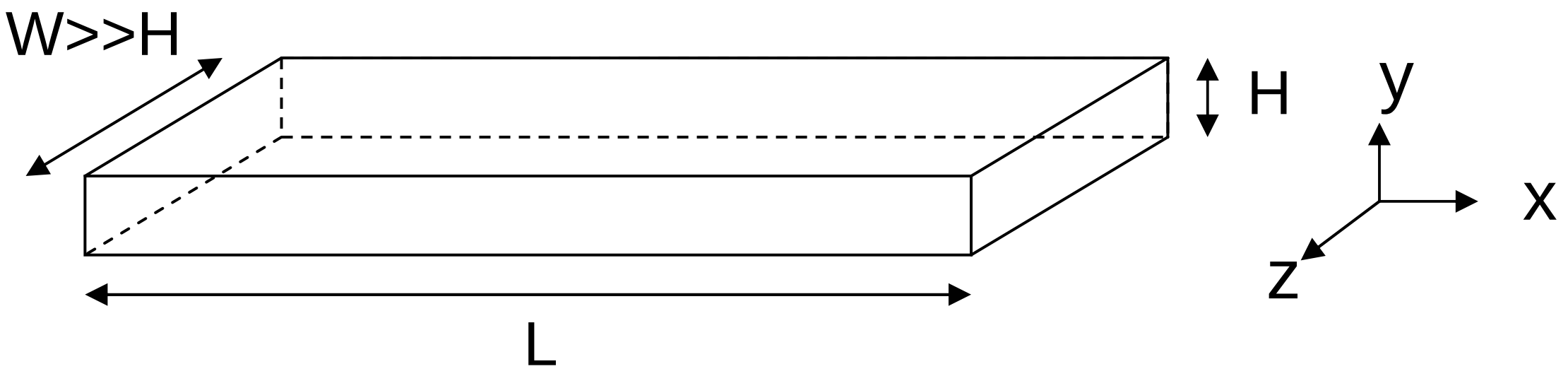

Fig.3.-

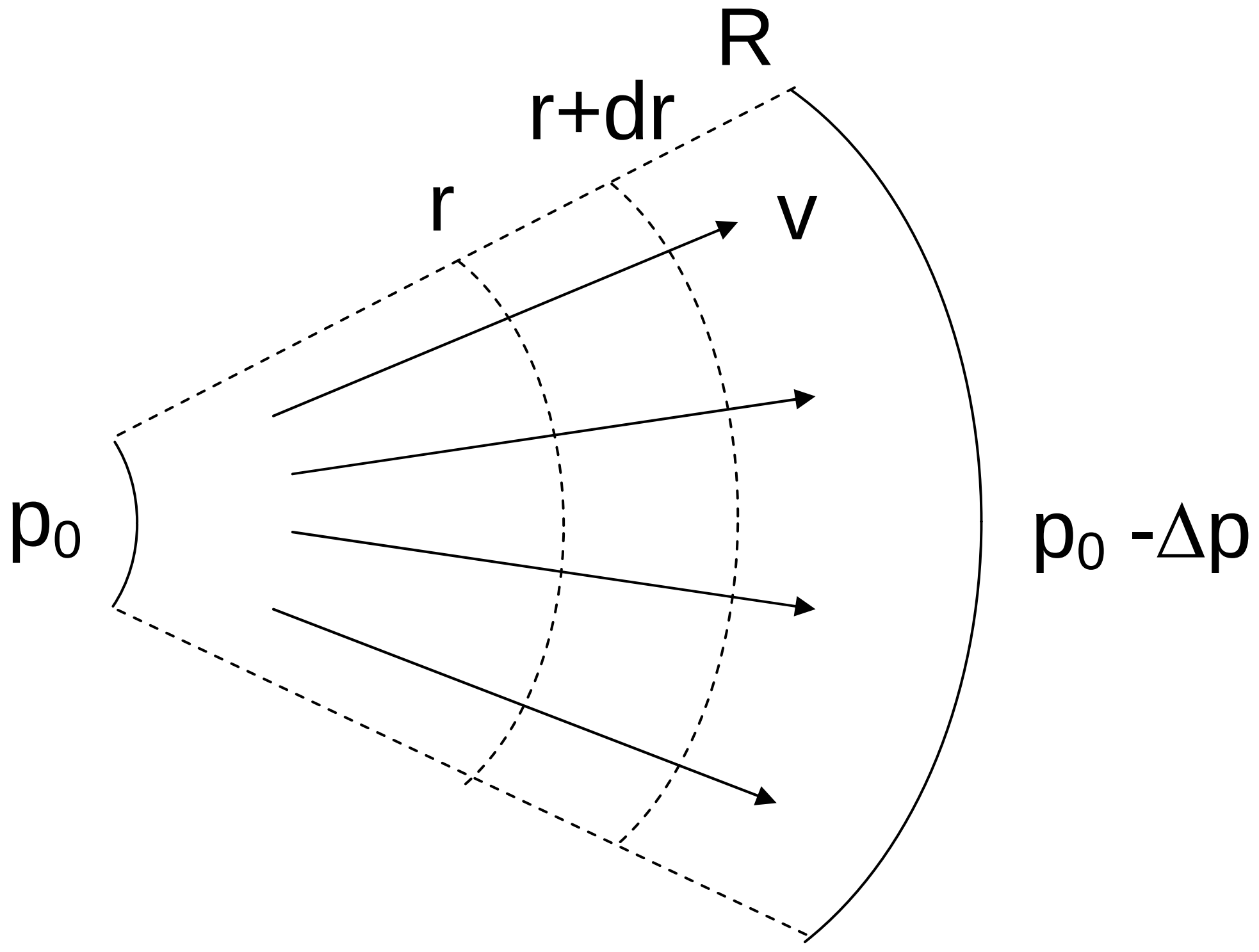

Fig.4.-

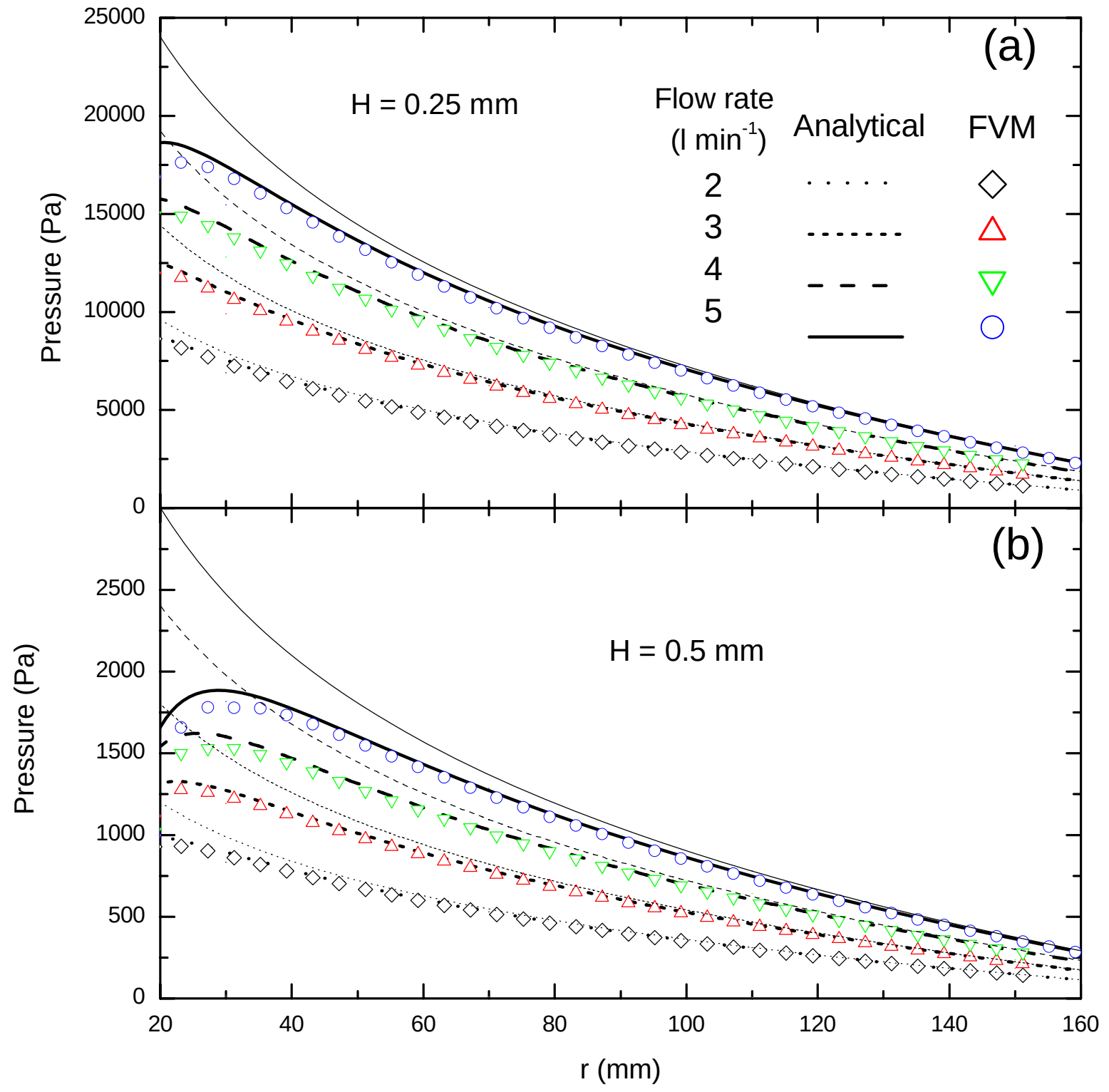

Figure 5

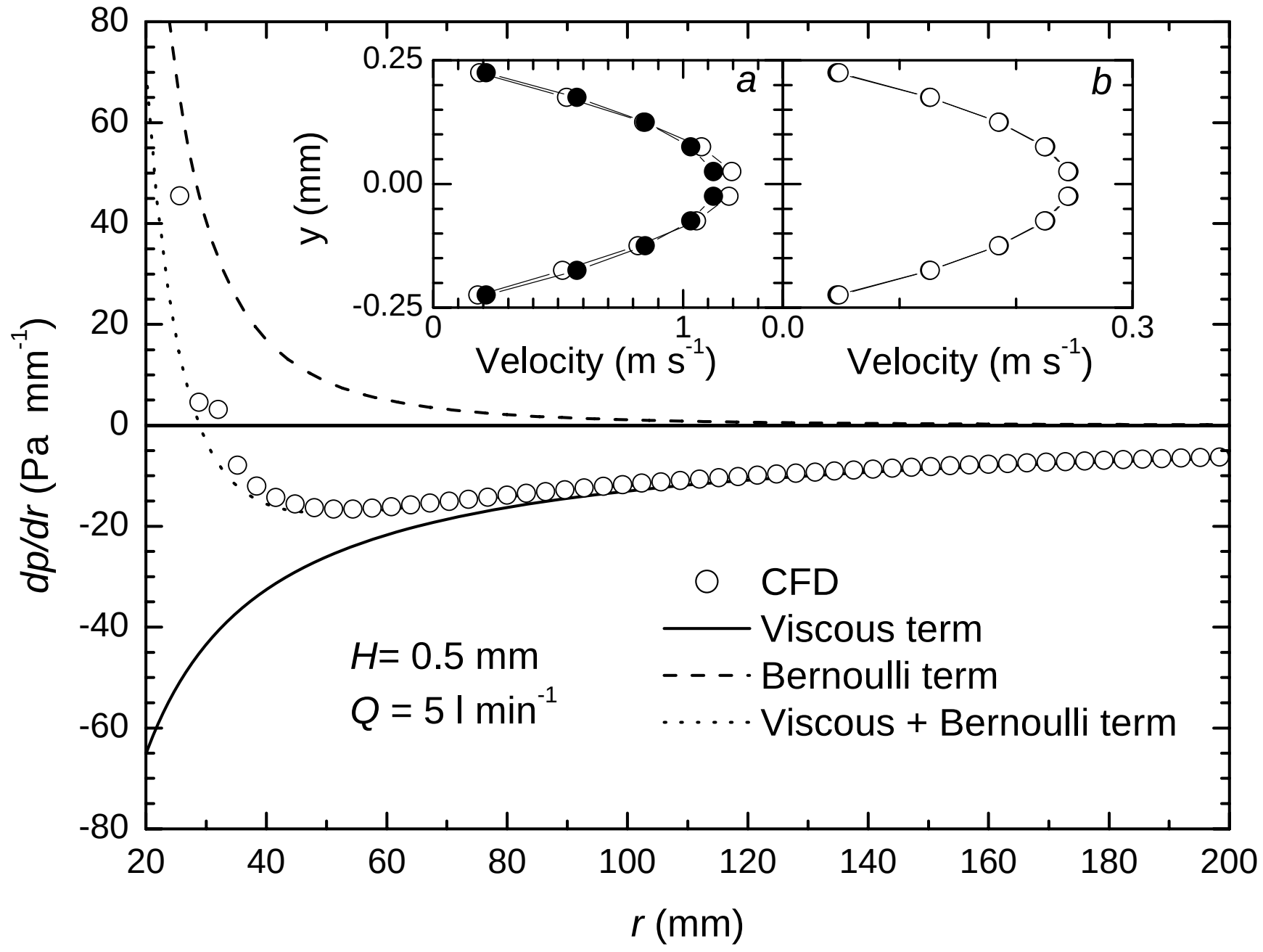

Fig.6.-

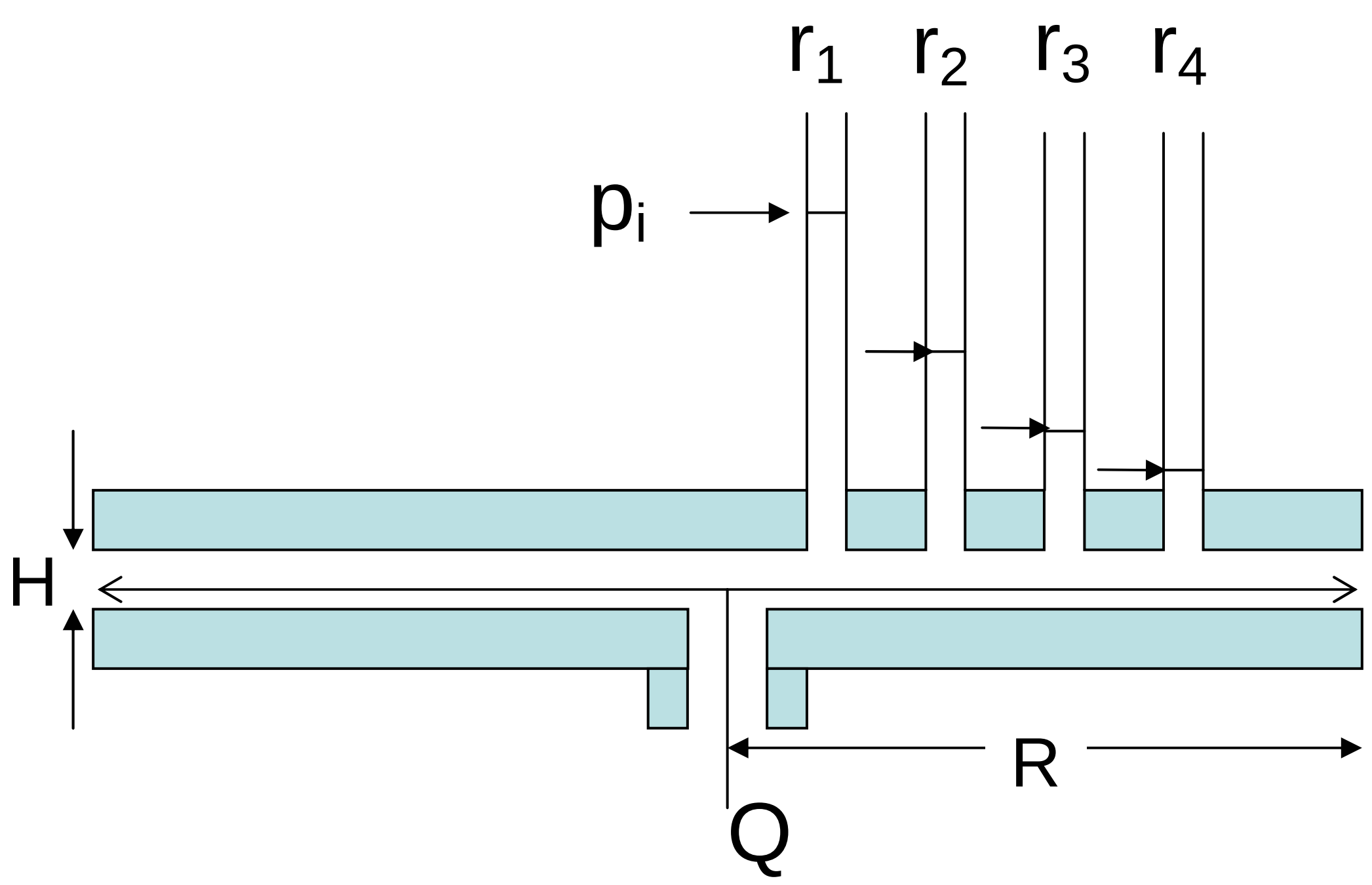

Fig.7.-

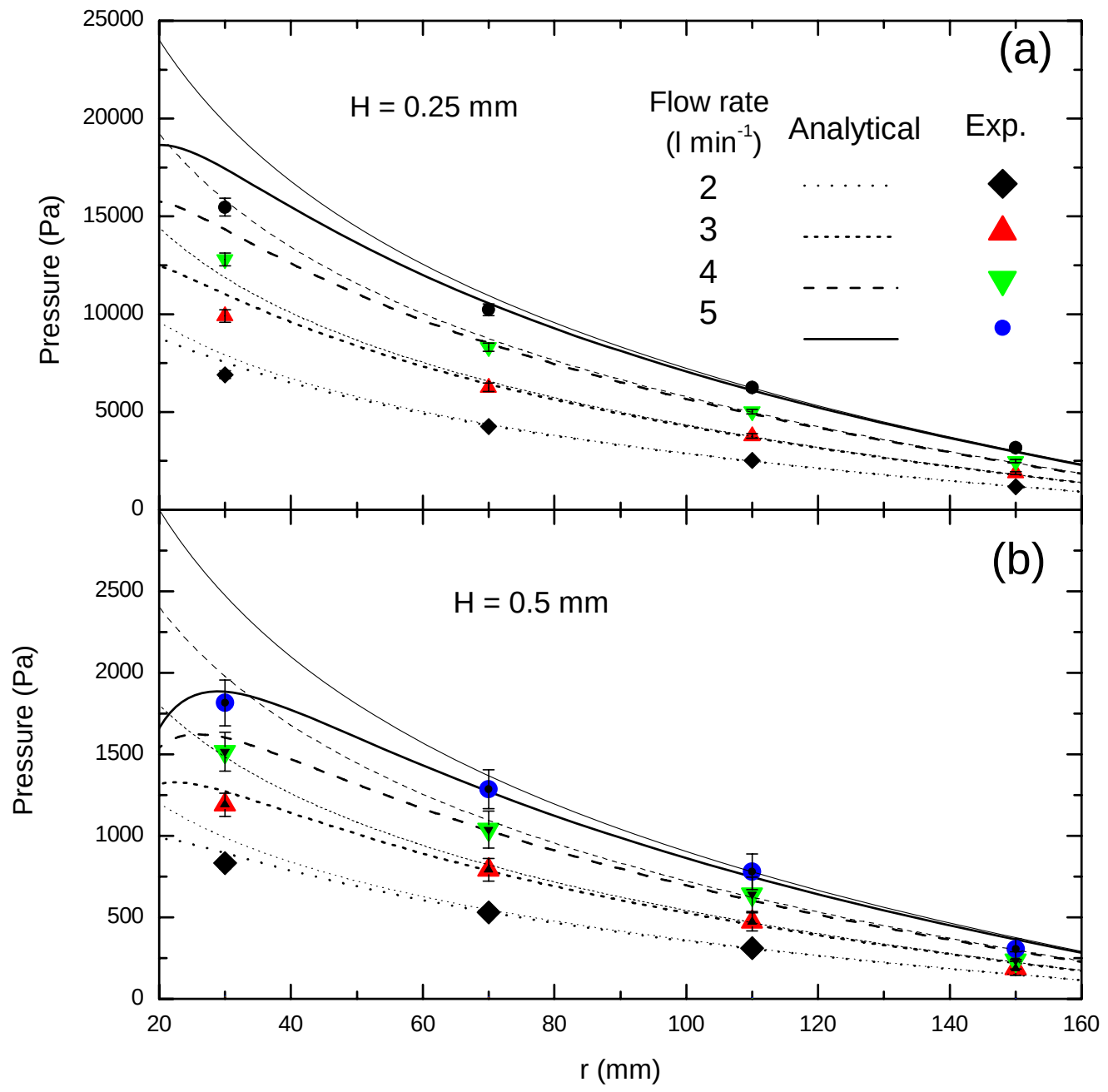

Figure 8.

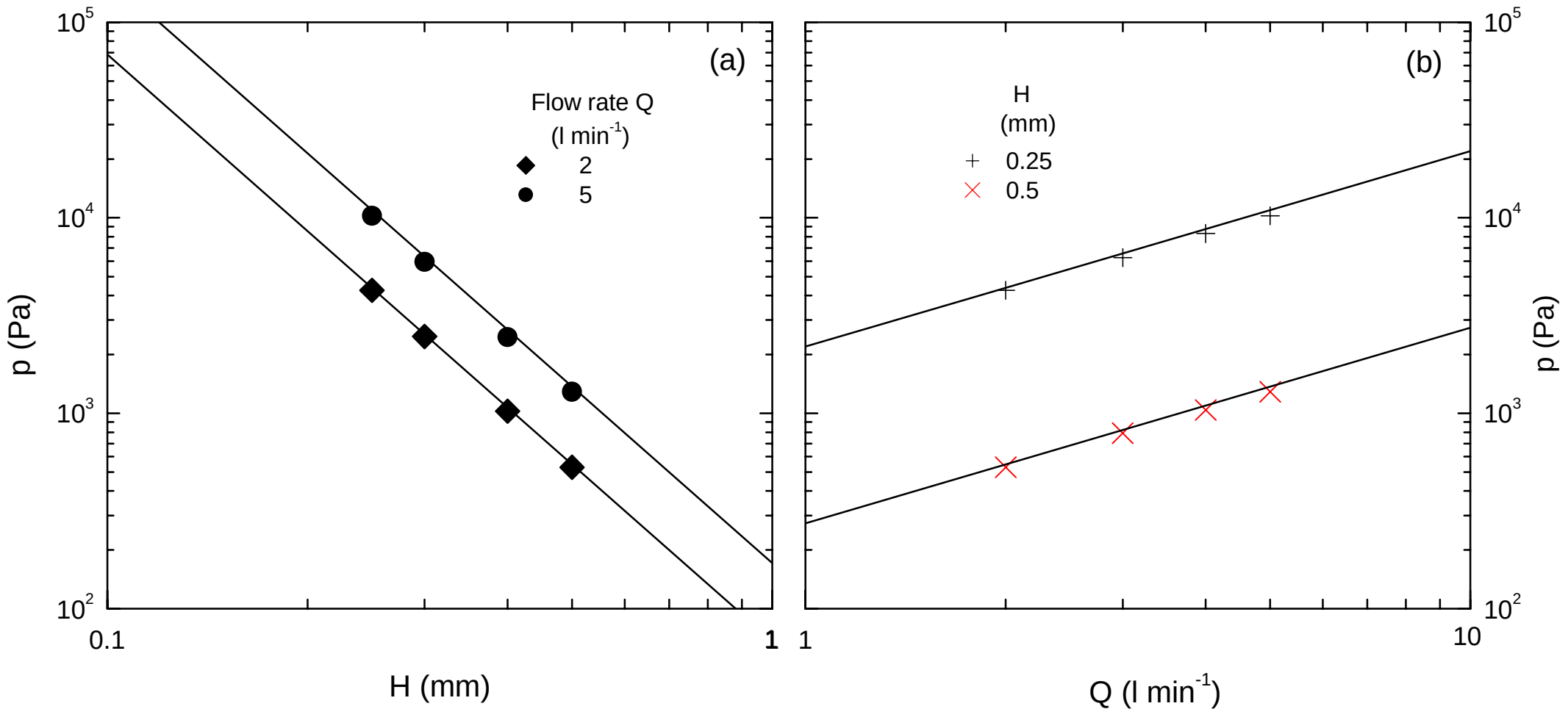

Figure 9